\documentclass[%
superscriptaddress,
preprint,
amsmath,amssymb,
]{revtex4-2}

\usepackage{multirow}
\usepackage{graphicx}
\usepackage{dcolumn}
\usepackage{bm}
\usepackage{hyperref}

\begin{document}

\preprint{APS/123-QED}

\title{s-wave superconductivity in kagome metal CsV$_{3}$Sb$_{5}$ revealed by $^{121/123}$Sb NQR and $^{51}$V NMR measurements}

\author{Chao Mu}
\affiliation{Beijing National Laboratory for Condensed Matter Physics and Institute of Physics, Chinese Academy of Sciences, Beijing 100190, China}
\affiliation{School of Physical Sciences, University of Chinese Academy of Sciences, Beijing 100190, China}

\author{Qiangwei Yin}
\affiliation{Department of Physics and Beijing Key Laboratory of Opto-electronic Functional Materials $\&$ Micro-nano Devices, Renmin University of China, Beijing 100872, China}

\author{Zhijun Tu}
\affiliation{Department of Physics and Beijing Key Laboratory of Opto-electronic Functional Materials $\&$ Micro-nano Devices, Renmin University of China, Beijing 100872, China}

\author{Chunsheng Gong}
\affiliation{Department of Physics and Beijing Key Laboratory of Opto-electronic Functional Materials $\&$ Micro-nano Devices, Renmin University of China, Beijing 100872, China}

\author{Hechang Lei}
\affiliation{Department of Physics and Beijing Key Laboratory of Opto-electronic Functional Materials $\&$ Micro-nano Devices, Renmin University of China, Beijing 100872, China}


\author{Zheng Li}
\email{lizheng@iphy.ac.cn}
\affiliation{Beijing National Laboratory for Condensed Matter Physics and Institute of Physics, Chinese Academy of Sciences, Beijing 100190, China}
\affiliation{School of Physical Sciences, University of Chinese Academy of Sciences, Beijing 100190, China}

\author{Jianlin Luo}
\email{jlluo@iphy.ac.cn}
\affiliation{Beijing National Laboratory for Condensed Matter Physics and Institute of Physics, Chinese Academy of Sciences, Beijing 100190, China}
\affiliation{School of Physical Sciences, University of Chinese Academy of Sciences, Beijing 100190, China}
\affiliation{Songshan Lake Materials Laboratory, Dongguan 523808, China}


\begin{abstract}
We report $^{121/123}$Sb nuclear quadrupole resonance (NQR) and $^{51}$V nuclear magnetic resonance (NMR) measurements on kagome metal CsV$_3$Sb$_5$ with $T_{\rm c}=2.5$ K. Both $^{51}$V NMR spectra and $^{121/123}$Sb NQR spectra split after a charge density wave (CDW) transition, which demonstrates a commensurate CDW state. The coexistence of the high temperature phase and the CDW phase between $91$ K and $94$ K manifests that it is a first order phase transition. At low temperature, electric-field-gradient fluctuations diminish and magnetic fluctuations become dominant. Superconductivity emerges in the charge order state. Knight shift decreases and $1/T_{1}T$ shows a Hebel--Slichter coherence peak just below $T_{\rm c}$, indicating that CsV$_3$Sb$_5$ is an s-wave superconductor.
\end{abstract}


\maketitle


Kagome lattice has been intensely studied in condensed matter physics due to its geometric frustration and nontrivial band topology. Depending on the electron filling, on-site repulsion $U$, and nearest-neighbor Coulomb interaction
$V$, several possible states have been
proposed as the ground state\cite{Wang2013Competing,Kiesel2013Unconventional}, such as quantum spin liquid\cite{Yan2011Spin,Fu2015NMR,Khuntia2020Gapless}, charge density wave (CDW)\cite{Guo2009Topological}, and superconductivity\cite{WenXG2009,Yu2012Chiral}.

Recently, superconductivity was discovered in the kagome metal $A$V$_3$Sb$_5$ ($A$\, =\, K, Rb, Cs), which undergos a CDW transition before the superconducting transition\cite{Ortiz2019New,Ortiz2020Cs,Ortiz2021KVSb,Lei2021Rb}.
Scanning tunneling microscopy (STM) observed a chiral charge order that breaks time-reversal symmetry\cite{jiang2020discovery} and leads to the anomalous Hall effect in the absence of magnetic local moments\cite{Yang2020Hall,Kenney2021muon,yu2021concurrence,feng2021chiral}. This chiral charge order is energetically preferred and tends to yield orbital current\cite{denner2021analysis}.
Inelastic x-ray scattering and Raman scattering exclude strong electron-phonon coupling driven CDW\cite{li2021observation}, while optical spectroscopy supports that CDW is driven by nesting of Fermi surfaces\cite{zhou2021origin}.
High pressure can suppress CDW transition and reveal a superconducting dome in the $P$--$T$ phase diagram\cite{chen2021double,du2021pressuretuned}. Further increasing pressure, a new superconducting phase emerges\cite{zhang2021pressureinduced,chenXL2021highly,zhao2021nodal}.

STM studies observed possible Majorana modes\cite{liang2021threedimensional} and pair density wave\cite{chen2021roton}, which points to unconventional superconductivity in the surface state. Signatures of spin-triplet pairing and an edge supercurrent have been observed in Nd/K$_{1-x}$V$_3$Sb$_5$ devices\cite{wang2020proximityinduced}. Whether the exotic surface superconducting state is intrinsic or comes from heterostructures, such as Bi$_2$Te$_3$/NbSe$_2$\cite{Xu2015BiTe}, needs to clarify the gap symmetry in the bulk state first. Thermal conductivity measurements on CsV$_3$Sb$_5$ down to $0.15$ K showed a finite residual linear term, which suggests a nodal gap and points to unconventional superconductivity\cite{zhao2021nodal}. On the other hand, magnetic penetration depth of CsV$_3$Sb$_5$ measured by tunneling diode oscillator showed a clear exponential behavior at low temperatures, which provides an evidence for nodeless superconductivity but does not rule out fully--gapped unconventional superconductivity\cite{duan2021nodeless}.

In this work, we report nuclear magnetic resonance (NMR)
and nuclear quadrupole resonance (NQR) investigations on CsV$_{3}$Sb$_{5}$. The splitting of $^{51}$V spectra and $^{121/123}$Sb spectra demonstrates that a commensurate CDW order forms at $94$ K with a first-order transition and coexists with superconductivity.
In the superconducting state, Knight shift decreases and a Hebel--Slichter coherence peak appears in the spin-lattice relaxation rate just below $T_{\rm c}$, which provides evidences for s--wave superconductivity in CsV$_{3}$Sb$_{5}$.

Single crystals of CsV$_{3}$Sb$_{5}$ were synthesized using the self-flux method, as reported in Ref.\cite{chen2021double}. Superconductivity with $T_{\rm c} =2.5$ K is confirmed by dc magnetization measured using a superconducting quantum interference device (SQUID).
The NMR and NQR measurements were performed using a phase coherent spectrometer. The spectra were obtained by frequency step and sum method which sums the Fourier transformed spectra at a series of frequencies\cite{Clark1995FSS}. The spin-lattice relaxation time $T_{1}$ was measured using a single saturation pulse.

\begin{figure*}
\includegraphics[width=0.98\textwidth,clip]{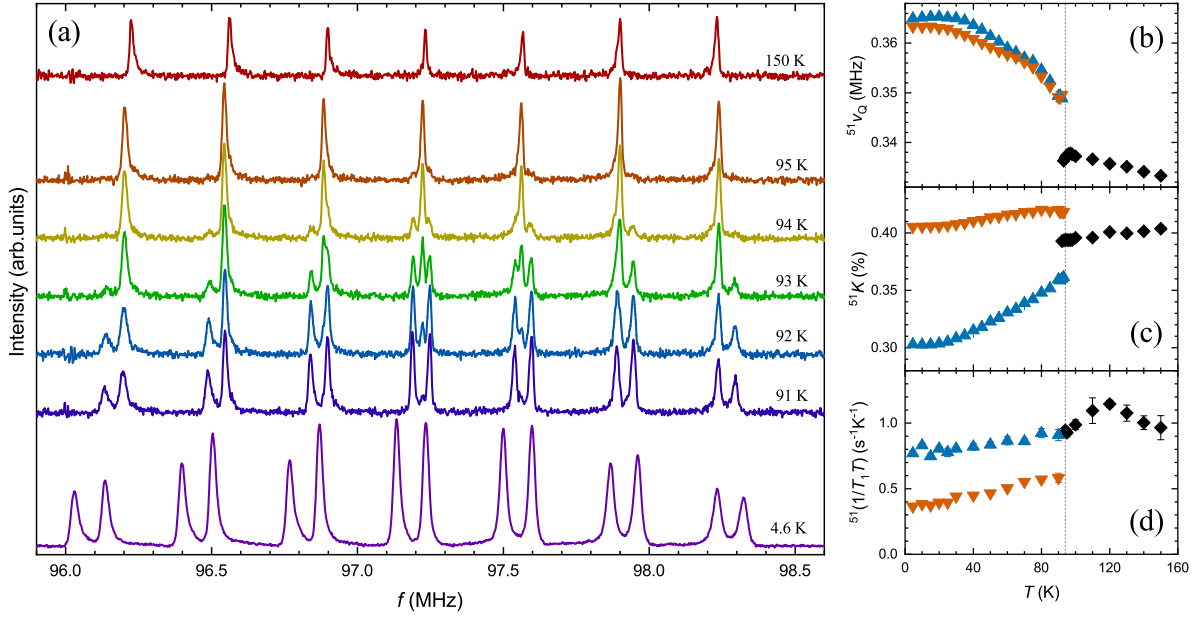}
\caption{\label{fig:V}(a) $^{51}$V-NMR spectra with $H \parallel c$. (b) Temperature dependence of quadrupole resonance frequency $\nu_{\rm Q}$. (c) Temperature dependence of the Knight shift of $^{51}$V. (d)Temperature dependence of the $^{51}(1/T_{1}T)$ measured at the central peaks. The vertical dashed line indicates the position of $T_{\rm CDW}$.}
\end{figure*}

Figure \ref{fig:V} (a) shows the spectra of $^{51}$V with $\mu _{0} H~(=8.65$ T$) \parallel c$. $^{51}$V with $I=7/2$ has seven NMR peaks at $150$ K. Below $T_{\rm CDW}=94$ K, the magnitude of original peaks decreases gradually and two sets of new peaks arise, which indicates that the CDW transition is commensurate and there are two different sites of V atoms in the CDW state. The coexistence of two phases between $91$ K and $94$ K demonstrates that the CDW transition is of first-order due to a simultaneous superlattice transition\cite{Ortiz2020Cs}. The quadrupole resonance frequencies of both $^{51}$V sites increase with decreasing temperature below $94$ K, as shown in Fig. \ref{fig:V} (b).

Knight shift $^{51}K$ is deduced from the central peaks and is summarized in Fig. \ref{fig:V} (c). Below $T_{\rm CDW}$, two splitting peaks give two Knight shift values, one of which jumps up and the other jumps down. The corresponding $^{51}(1/T_{1}T)$ jumps to the opposite side or does not change at $T_{\rm CDW}$, as shown in Fig. \ref{fig:V} (d). We also measured Knight shift and $^{51}(1/T_{1}T)$ the with $H \perp c$, as shown in  Fig S$1$ of the Supplemental Material, which also shows an opposite trend at $T_{\rm CDW}$. Therefore, the splitting of $^{51}K$ is not caused by different hyperfine couplings, but by the change of the orbital part of Knight shift.

$^{51}(1/T_{1}T)$ in a Fermi liquid state is proportional to the square of the density of states (DOS) around the Fermi energy $N(E_{\rm F}$). Just below $T_{\rm CDW}$, $^{51}(1/T_{1}T)$ of the right peak jumps to $64 \%$ of the value above $T_{\rm CDW}$, indicative of the decrease of $N(E_{\rm F})$ by $20 \%$. $^{51}(1/T_{1}T)$ of the left peak is smooth at $T_{\rm CDW}$. The difference of $^{51}(1/T_{1}T)$ between two V sites manifests the modulation of DOS at different V positions, which is consistent with $2 \times 2$ CDW charge order\cite{chen2021roton}. 
$^{51}$V has a very small quadrupole moment, $Q = -5.2\times 10^{-30}$ m$^{2}$. Therefore, the relaxation is governed by magnetic fluctuations and can reflect the change in the DOS, even though the electric field gradient (EFG) fluctuates strongly near $T_{\rm CDW}$. On the contrary, $^{121}$Sb and $^{123}$Sb have larger quadrupole moments, $^{121}Q = -54.3\times 10^{-30}$ m$^{2}$ and $^{123}Q = -69.2 \times 10^{-30}$ m$^{2}$ respectively, which can be used to detect the EFG fluctuations and perform NQR measurement.

\begin{figure*}
\includegraphics[width=0.98\textwidth,clip]{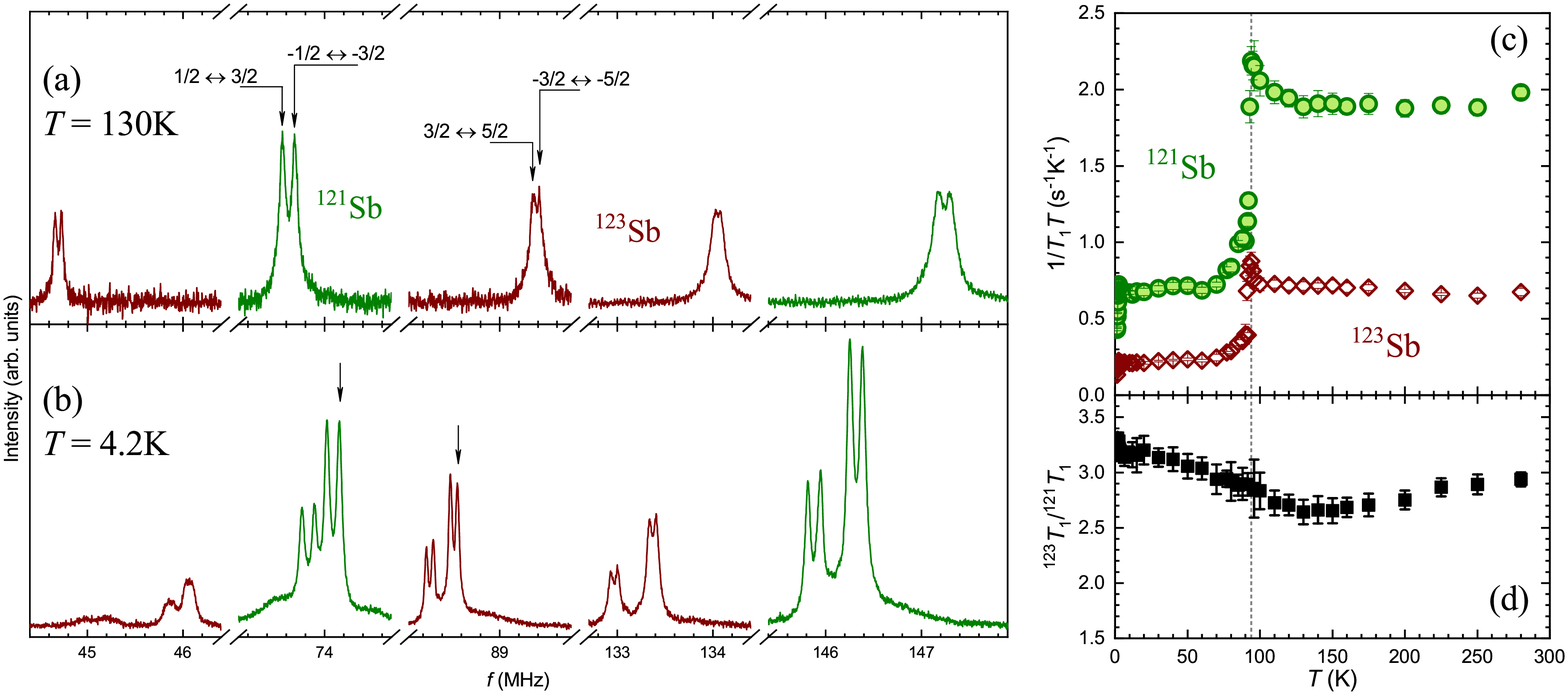}
\caption{\label{fig:Sb} $^{121/123}$Sb-NQR spectra with a perturbing field of $64$ Oe along $c$-axis at (a) $130$ K and (b) $4.2$ K, respectively. The arrows mark the peaks where $T_{1}$ was measured. (c) Temperature dependence of the $1/T_{1}T$ measured with $^{121}$Sb at $-1/2\leftrightarrow -3/2$ peak and $^{123}$Sb at $-3/2\leftrightarrow -5/2$ peak, respectively. (d) Temperature dependence of the $^{121/123}$Sb isotopic ratio $^{123}T_{1}/^{121}T_{1}$. The vertical dashed line indicates the position of $T_{\rm CDW}$.}
\end{figure*}

In order to study the superconducting state, we must perform measurement at a magnetic field lower than the upper critical field, $\mu_{0} H_{\rm c2} \sim 0.5$\, T\cite{zhao2021nodal}. Usually NQR experiments do not need applying magnetic field, however Knight shift cannot be measured at zero field. Here we measure $^{121}K$ by applying a perturbing field of $64$ Oe that is larger than the lower critical field, $\mu_{0} H_{\rm c1} \sim 2$\,mT\cite{Yuli2021}. With a perturbing field, the nuclear spin Hamiltonian is the sum of electric quadrupole interactions and the nuclear Zeeman energy\cite{Sakurai2008}
\begin{equation} \label{eq:vQ} 
\begin{aligned}
\mathcal{H} = &\mathcal{H}_{Q} +\mathcal{H}_{Z}\\
= &\frac{h\nu_{zz}}{6}\left[ (3I_{z}^{2}-I^{2})+\frac{\eta (I_{+}^{2}+I_{-}^{2})}{2}  \right] \\
& - \gamma\hbar(1+K)H_{0} \left( I_{z} \cos \theta + \frac{I_{+} e^{-i \phi}+I_{-} e^{i \phi} }{2}\sin \theta\right)
\end{aligned}
\end{equation}
where $\nu _{zz}$ is the quadrupole resonance frequency along the principal axis ($c$-axis), $\nu _{zz} \equiv \frac{3e^{2}qQ}{2I(2I-1)}$, with $eq=V_{zz}$. $\eta$ is an asymmetry parameter of the EFG, $\eta=\frac{V_{xx}-V_{yy}}{V_{zz}}$. $V_{xx}$, $V_{yy}$, $V_{zz}$ are the EFGs along the $x$, $y$, $z$ directions respectively. $\theta$ and $\phi$ are the polar and azimuthal angles between the direction of the applied field and the principal axis of EFG. When $\eta$ is small enough to be negligible, the resonance frequency at a small perturbation field along $c$-axis is
\begin{equation} \label{eq:vQ}  
\nu_{\rm NQR}(H_{0}) = \nu_{\rm NQR}(0) \mp \gamma (1+K) H_{0}, \ I_{ z}=\pm|I_{ z}|
\end{equation}
The perturbing field eliminates the degeneration of $\pm I_{z}$ and results in a spectra splitting. When $H_{0} \perp c$, the peak of $|{\pm}\frac{1}{2}\rangle \leftrightarrow |{\pm}\frac{3}{2}\rangle$ is split by the perturbing field, while other peaks are not affected\cite{Asayama2002}.
\begin{equation} \label{eq:vQa} 
\begin{aligned}
\nu_{\rm NQR}(H_{0}) =& \nu_{\rm NQR}(0) \pm \frac{I+\frac{1}{2}}{2}\gamma (1+K) H_{0}, \ |{\pm}\frac{1}{2}\rangle \leftrightarrow |{\pm}\frac{3}{2}\rangle \\
\nu_{\rm NQR}(H_{0}) =& \nu_{\rm NQR}(0), \ |{\pm}m\rangle \leftrightarrow |{\pm}(m+1)\rangle, \ m\geq\frac{3}{2}
\end{aligned}
\end{equation}
where $m$ is a half-integer.

There are two isotopes of Sb nuclei which are $^{121}$Sb with $I=5/2$ and $^{123}$Sb with $I=7/2$. The NQR spectra should contain two $^{121}$Sb peaks and three $^{123}$Sb peaks. Figures \ref{fig:Sb} (a) and \ref{fig:Sb}(b) show the spectra with a perturbing field of $64$ Oe along $c$-axis, which eliminates the degeneration of $\pm I_{z}$ and results in a small spectral splitting. There are two different crystallographic sites for Sb in CsV$_{3}$Sb$_{5}$ with the atomic ratio of Sb$1:$  Sb$2=1:4$\cite{Ortiz2019New}. The relative intensities of different sites are proportional to their atomic ratio\cite{Julien2013}, so Sb$2$ site has larger spectral intensity than Sb$1$ site. The spectra in Figs. \ref{fig:Sb} (a) and \ref{fig:Sb} (b) are of Sb$2$ atoms which encapsulate the kagome layer with a graphite-like network. Sb$1$ atoms located in the kagome plane with V atoms have weak spectra, which are presented in the Supplemental Material. The $\nu _{zz}$ and $\eta$ deduced from the spectra above and below $T_{\rm CDW}$ are summarized in Table \ref{tab:Sb}, where $\eta$ is zero above $T_{\rm CDW}$ reflecting isotropy in the $xy$-plane. Below $T_{\rm CDW}$, $\eta$ of Sb$2$ changes to finite values, which indicate the breaking of in-plane symmetry and two unequal sites of Sb$2$ with the ratio of $1:2$.  The slightly decreasing $\nu _{zz}$ implies that the main change is not the EFG strength, but the EFG direction at Sb$2$ sites. On the other hand, the EFG at Sb$1$ site changes frequency obviously without asymmetry.

\begin{table}[b]
\caption{\label{tab:Sb}%
The quadrupole resonance frequencies and asymmetry parameters.
}
\begin{ruledtabular}
\begin{tabular}{clccc}
\textrm{$T$}&
\textrm{site}&
\textrm{$\eta$}&
\textrm{$^{121}\nu _{zz}$}&
\textrm{$^{123}\nu _{zz}$}\\
\colrule
\multirow{2}{*}{130 K} & Sb$1$ & 0 & 71.41 & 43.35\\
& Sb$2$ & 0 & 73.62 & 44.69\\
\colrule
\multirow{3}{*}{4.2 K} & Sb$1$ & 0 & 78.19 & 47.16\\
 & Sb$2'$ & 0.097 & 73.08 & 44.36\\
 & Sb$2''$ & 0.099 & 73.31 & 44.50\\
\end{tabular}
\end{ruledtabular}
\end{table}

$T_{1}$ was measured at $|\textnormal{-}\frac{1}{2}\rangle \leftrightarrow |\textnormal{-}\frac{3}{2}\rangle$ peak of $^{121}$Sb$2$  and $|\textnormal{-}\frac{3}{2}\rangle \leftrightarrow |\textnormal{-}\frac{5}{2}\rangle$ peak of $^{123}$Sb$2$ respectively. After spectra splitting below $T_{\rm CDW}$, $T_{1}$ was measured at the right-side peaks, marked by arrows in Fig. \ref{fig:Sb} (b). The degeneration is eliminated by the magnetic field, so the NQR recovery curves are no longer applicable. The recovery curve used for $|\textnormal{-}\frac{1}{2} \rangle \leftrightarrow |\textnormal{-}\frac{3}{2}\rangle$ transition of $^{121}$Sb is
\begin{equation}\label{eq:RC52} 
\begin{aligned}
1 - \frac{M(t)}{M(\infty)} =& \frac{1}{35} e^{-\frac{t}{T_1}}
+ \frac{3}{56} e^{-\frac{3t}{T_1}}
+ \frac{1}{40}e^{-\frac{6t}{T_1}} \\
& +  \frac{25}{56} e^{-\frac{10t}{T_1}}
+ \frac{25}{56} e^{-\frac{15t}{T_1}}
\end{aligned}
\end{equation}
The recovery curve used for $|\textnormal{-}\frac{3}{2} \rangle \leftrightarrow |\textnormal{-}\frac{5}{2} \rangle$ transition of $^{123}$Sb is
\begin{equation}\label{eq:RC73} 
\begin{aligned}
1 - \frac{M(t)}{M(\infty)}  = & \frac{1}{84} e^{-\frac{t}{T_1}}
+ \frac{1}{21}e^{-\frac{3t}{T_1}}
+ \frac{1}{132}e^{-\frac{6t}{T_1}}
+ \frac{25}{308}e^{-\frac{10t}{T_1}} \\
&+ \frac{100}{273}e^{-\frac{15t}{T_1}}
+\frac{49}{132}e^{-\frac{21t}{T_1}}
+ \frac{49}{429}e^{-\frac{28t}{T_1}}
\end{aligned}
\end{equation}
We use the same fitting function for both above and below $T_{\rm CDW}$, since the $\eta$ is small. Figure \ref{fig:Sb} (c) shows the temperature dependence of $1/T_{1}T$ of $^{121/123}$Sb. Upon cooling, $^{121/123}(1/T_{1}T)$ starts to increase a little and drops sharply below $T_{\rm CDW}$ for both $^{121}$Sb and $^{123}$Sb. With further cooling, $^{121/123}(1/T_{1}T)$ becomes constant from $70$ K to $T_{\rm c}$, which obeys the Korringa relation and suggests that the system is in a Fermi liquid state. $^{51}(1/T_{1}T)$ of $^{51}$V has shown that DOS loss is only $20\%$. Here the large drop implies that the EFG fluctuations play an important role above $T_{\rm CDW}$ and are suppressed after CDW transition. Moreover, $^{121/123}(1/T_{1}T)$ decreases continuously without sudden jump like $^{51}(1/T_{1}T)$. This implies that the superlattice transition has strong effect on V atoms that locate in the kagome plane, but has very small effect on Sb$2$ atoms that locate outside the kagome plane. Therefore, $^{121/123}(1/T_{1}T)$ of Sb$2$ shows a second--order CDW transition behavior.

In general, a spin-lattice relaxation occurs through magnetic and/or electric-quadrupole channels. The magnetic relaxation is related to the gyromagnetic ratio $\gamma _{n}$  and gives $^{123}T_{1}/^{121}T_{1}=3.4$, while the electric-quadrupole relaxation is related to the quadruple moment $Q$ and gives $^{123}T_{1}/^{121}T_{1}=1.5$\cite{Ishida2013}. Above $T_{\rm CDW}$, $^{123}T_{1}/^{121}T_{1}$ is smaller than $3$, as shown in Fig. \ref{fig:Sb} (d), which indicates that both magnetic fluctuations and EFG fluctuations are important. Below $T_{\rm CDW}$, the ratio increases with decreasing temperature and approaches to $3.4$ at zero temperature limit, which indicates that the magnetic process becomes dominant at low temperature. Therefore, the effect of EFG fluctuations in the superconducting state is negligible.

\begin{figure}
\includegraphics[width=0.49\textwidth,clip]{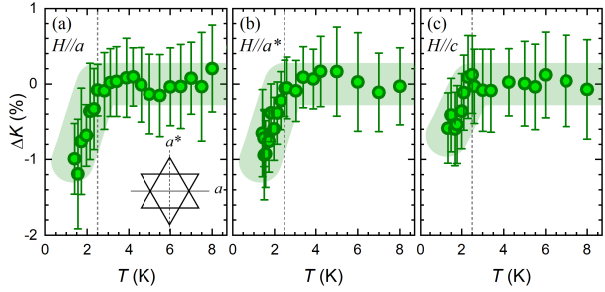}
\caption{\label{fig:SbK} Temperature dependence of $\Delta K$ of $^{121}$Sb with (a) $H \parallel a$ (b) $H \parallel a^{*}$ and (c) $H \parallel c$, where $a$ and $a^{*}$ are orthogonal directions in the basal plane. The vertical dashed lines indicate the position of $T_{\rm c}$}
\end{figure}

\begin{figure}[htb]
\includegraphics[width=0.45\textwidth,clip]{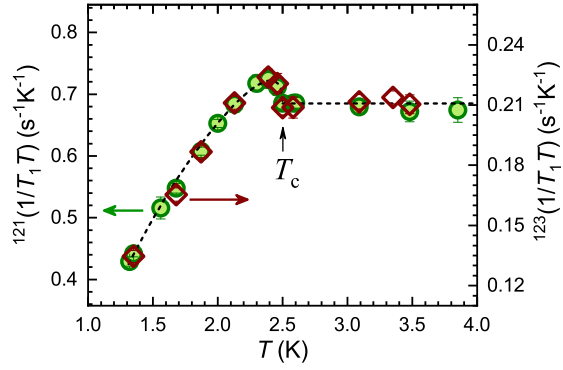}
\caption{\label{fig:SbT1T} Temperature dependence of $^{121}(1/T_{1}T)$ (left axis) and $^{123}(1/T_{1}T)$ (right axis). A Hebel-Slichter coherence peak appears just below $T_{\rm c}$. The curve and line are guides to the eyes.}
\end{figure}

The Knight shift of Sb can be deduced from the spectra splitting. The field calibrated by a Gauss meter has an uncertainty, so the absolute Knight shift value is not reliable in an ultra-low field. The relative Knight shift $\Delta K=K-K_{0}$, where $K_{0}$ is the average value above $T_{\rm c}$, has very weak dependence on field. The $\Delta K$ values for all directions are almost unchanged above $T_{\rm c}$ and decrease below $T_{\rm c}$, as shown in Fig. \ref{fig:SbK}, which is consistent with spin-singlet pairing. The perturbing field is much larger than the lower critical field, so the drop below $T_{\rm c}$ is not from Meissner effect but intrinsic spin susceptibility.

In an ultra-low field, $^{121}K$ measurements have large errors which preclude further analysis. On the other hand, spin-lattice relaxation rate can be measured accurately. $^{121/123}(1/T_{1}T)$ shows a clear Hebel--Slichter coherence peak just below $T_{\rm c}$ and then rapidly decreases at low temperatures as shown in Fig. \ref{fig:SbT1T}. It evidences that the superconducting gap is of s-wave symmetry.
As for d or p-wave, gap sign changes over the Fermi surface, which suppresses a coherence peak. In iron-based superconductors, s$^{\pm}$-wave also changes gap sign between different Fermi surfaces and suppresses a coherence peak\cite{Parker2008}. So far, the Hebel-Slichter coherence peak has only been observed in s-wave superconductors.
Here the coherence peak is small, which may be due to residual EFG fluctuations\cite{Li2016TaPdTe} and/or a strong electron-phonon coupling\cite{Kotegawa2001MgB2}. In other s-wave superconductors, the coherence peaks of Sb are also very small, such as in YbSb$_2$\cite{KOHORI2003} and BaTi$_{2}$Sb$_{2}$O\cite{Ishida2013}. A recent STM study has suggested the absence of sign-change in the superconducting order parameter\cite{xu2021multiband}. Magnetic penetration depth measured by tunneling diode oscillator has shown a nodeless behavior\cite{duan2021nodeless}. These results are both consistent with the s-wave gap symmetry. There are also experimentally observed residual density of states in the superconducting state\cite{zhao2021nodal,xu2021multiband}, which may be due to the competition between superconductivity and CDW. Another s-wave superconductor Ta$_{4}$Pd$_{3}$Te$_{16}$ also has a CDW order coexisting with superconductivity and shows residual density of states in the superconducting state\cite{Pan2015TPT}.

$^{121}(1/T_{1}T)$ is measured at the split peak due to the CDW transition and shows sudden changes at both $T_{\rm CDW}$ and $T_{\rm c}$, which indicates that the superconductivity coexists with the CDW state.
The coexistence of superconductivity and topological charge order may stimulate exotic excitations in the surface state\cite{liang2021threedimensional,Fu2008}, which needs further investigations. Recently another superconducting phase at high pressure was found in CsV$_{3}$Sb$_{5}$\cite{chenXL2021highly,zhao2021nodal}. Whether the high-pressure superconducting phase is the same as that at ambient pressure needs further studies.

In conclusion, we have performed $^{51}$V NMR and $^{121/123}$Sb NQR studies on CsV$_{3}$Sb$_{5}$. The variation of the NMR and NQR spectra below $T_{\rm CDW}$ can be understood by the occurrence of  a first-order commensurate CDW transition. EFG fluctuations are suppressed and magnetic fluctuations become dominant at low temperature. In the superconducting state, $^{121}K$ decreases and $^{121/123}(1/T_{1}T)$ shows a coherence peak just below $T_{\rm c}$, which is the hallmark of the s-wave full gap superconductivity. The superconductivity coexists with the CDW state, which may lead to exotic surface state.

We thank S. K. Su, X. B. Zhou and B. Su for assistance in some of the measurements.
This work was supported by the National Key Research and Development Program of China (Grant No. 2017YFA0302901, 2018YFE0202600, and 2016YFA0300504), the National Natural Science Foundation of China (Grant No. 11921004, 11634015, 11822412 and 11774423), the Beijing Natural Science Foundation (Grant No. Z200005), and the Strategic Priority Research Program and Key Research Program of Frontier Sciences of the Chinese Academy of Sciences (Grant No. XDB33010100).




\bibliography{CVS}

\end{document}